\documentclass[aip,apl, reprint]{revtex4-1}

\usepackage{amsmath}
\usepackage{epsfig,graphicx}
\usepackage[dvips,ps2pdf]{hyperref}
\usepackage{bm}

\begin{document}

\title{Superconducting microfabricated ion traps}

\author{Shannon X. Wang} \email[]{sxwang@mit.edu}
\author{Yufei Ge}
\author{Jaroslaw Labaziewicz}
\affiliation{Center for Ultracold Atoms, Research Laboratory of Electronics and Department of Physics, Massachusetts Institute of Technology, Cambridge, Massachusetts, 02139, USA}
\author{Eric Dauler}\affiliation{MIT Lincoln Laboratory, 244 Wood St., Lexington, Massachusetts, 02420, USA}
\author{Karl Berggren}
\author{Isaac L. Chuang}
\affiliation{Center for Ultracold Atoms, Research Laboratory of Electronics and Department of Physics, Massachusetts Institute of Technology, Cambridge, Massachusetts, 02139, USA}

\date{\today}

\begin{abstract}
\noindent We fabricate superconducting ion traps with niobium and niobium nitride and trap single $^{88}$Sr ions at cryogenic temperatures. The superconducting transition is verified and characterized by measuring the resistance and critical current using a 4-wire measurement on the trap structure, and observing change in the rf reflection. The lowest observed heating rate is 2.1(3) quanta/sec at 800~kHz at 6~K and shows no significant change across the superconducting transition, suggesting that anomalous heating is primarily caused by noise sources on the surface. This demonstration of superconducting ion traps opens up possibilities for integrating trapped ions and molecular ions with superconducting devices.
\end{abstract}

\maketitle

Microfabricated surface electrode ion traps have significantly advanced the capabilities of trapped ion systems for quantum information processing \cite{Chiaverini:05, Stick:06, Steane:07}, by enabling increased level of precision, density, and system integration. The ions trapped in these devices represent quantum bits and are confined by oscillating electric fields. While typical ion traps currently employ aluminum, gold, or doped semiconductor as the electrode material \cite{Seidelin:06, Leibrandt:09, Stick:06}, the anomalous electric field noise \cite{Turchette:00} affecting such traps provides significant motivation to explore qualitatively different materials for microfabricated ion traps, such as superconductors.  In particular, the fact that a superconductor expels electric fields provides an opportunity to test the theoretical understanding that anomalous noise results from surface patch potentials \cite{Turchette:00, Dubessy:09}, rather than sources in the bulk, since the bulk noise sources would be screened by the superconductor. A similar approach was taken for neutral atoms, where in superconducting traps it was found that magnetic near-field noise is suppressed resulting in lower heating rate and longer spin-flip lifetimes \cite{Fermani:10, Kasch:10}. 
For a thin film superconducting ion trap, blue lasers are typically employed for doppler cooling and state detection of trapped ions, and the short (279$-$422~nm \cite{James:98a}) wavelengths may create quasiparticles in the superconductor, driving it into a normal state. Therefore, verifying that the superconductor employed is actually superconducting during an experiment is required. 

Here, we demonstrate the operation of several superconducting microfabricated ion traps made with niobium and niobium nitride, describe how superconductivity is verified during trap operation, and apply these traps to test the physical mechanisms of anomalous noise. 
The demonstration of superconducting ion traps opens up possibilities for integrating trapped ions and molecular ions with superconducting devices, such as photon counting detectors, microwave resonators \cite{Schuster:09}, and circuit-QED systems \cite{Tian:04}.


The ion traps used in this experiment consist of Nb or NbN on a sapphire substrate. One Nb and one NbN trap are identical to a prior five-electrode design \cite{Labaziewicz:08}. An additional Nb trap (Nb-g) includes a thin wire structure \cite{Wang:09} on the center ground electrode that is electrically connected in a 4-wire configuration to measure the resistivity of the electrode. The thinnest part of the wire is 10~$\mu$m wide. The fabrication procedure is as follows. A 400~nm layer of Nb is grown by DC magnetron sputtering of a niobium target in Ar gas; NbN is grown by adding N$_2$ gas during sputtering. Electrodes are defined by optical lithography using NR9-3000P photoresist, exposed through a chrome mask and developed in RD6 developer. Reactive ion etch with CF$_4$ and O$_2$ is used to etch exposed metal. Gold contact pads for wirebonding are then defined by optical lithography using S1813 or NR9-3000PY photoresist, deposited using evaporation and created with a lift-off process. After the initial Nb sputtering, the trap is maintained below a temperature of 90$^\circ$C during all steps of the fabrication and packaging process to minimize oxide formation on the surface. For trap Nb-g, a surface-mount resistor (0603, 1~k$\Omega$) is glued to one trap corner and used as a heater for controlling the trap temperature. The trap is operated in a bath cryostat, and we estimate the trap surface temperature to be $\sim6$~K \cite{Antohi:09}. 

A single $^{88}$Sr ion is trapped 100~$\mu$m above the trap surface, loaded via photoionization of a thermal vapor. Typical ion lifetime is several hours with cooling lasers on, same as traps made of normal metals. Typical axial trap frequencies are 2$\pi \times$ 0.8$-$1.3~MHz. The 5S$_{1/2} \rightarrow $4D$_{5/2}$ optical transition is used for sideband cooling and temperature readout, addressed via a narrow 674~nm diode laser locked to a stable external cavity. 

\begin{figure}[tb]
\includegraphics[width=3.375in]{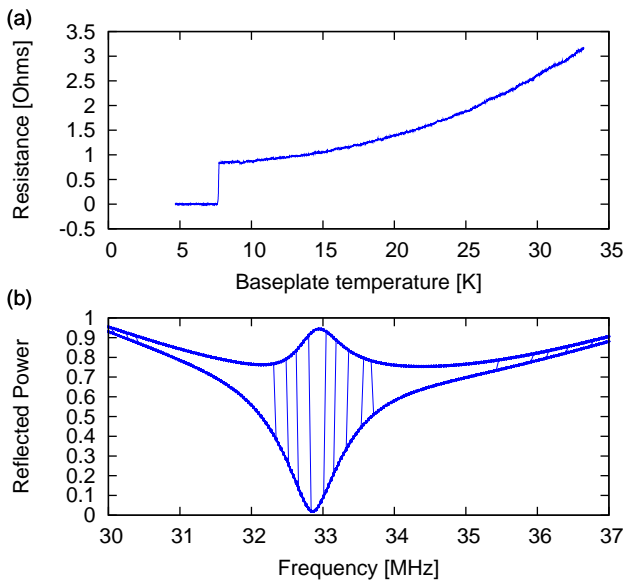}
\caption{\label{Fig:SCtrans} a) Resistance in Nb-g during warm-up of the cryostat. The baseplate temperature is lower than the trap temperature. b) Fraction of reflected rf power vs frequency in the NbN trap during warm-up of the cryostat; top curve: below $T_c$, bottom curve: above $T_c$. The observed value switches multiple times between the two curves due to warm-up and cool-down of the trap as the rf is moved on and off the resonant frequency. }
\end{figure}

We verify that the traps are superconducting by observing three variables: resistance, critical current, and reflected rf power. Resistance is measured on the wire structure in Nb-g as the trap cools or warms up. Figure \ref{Fig:SCtrans}a) shows the resistance as a function of measured baseplate temperature (with a Lakeshore RX103 calibrated diode) during a slow warm-up of the cryostat. The trap is heated to above $T_c$ during ion loading, but cools to below $T_c$ within 5-10 minutes. Superconductivity is maintained on the trap when 150~V (amplitude) of rf drive is applied to the trap rf electrodes to create the trapping potential. This corresponds to $\sim$250~mA of current on the rf electrodes, given a capacitance to ground of $\sim$8~pF. The critical current of the wire structure, both with and without the trapping lasers, is 180(1)~mA. This corresponds to a critical current density of 4$\times 10^6$~A/cm$^2$, typical in order-of-magnitude for thin-film Nb. Based on this measured critical current density and electrode dimensions (400~nm $\times$ 150$\mu$m), the calculated current limit on the rf electrodes is 2.7~A, well above what is needed for typical trap operations. The superconducting transition is also observed by looking at the reflected rf power in the NbN trap, which is more resistive immediately above $T_c$. Rf reflected power is measured with a directional coupler mounted before the helical resonator, which is inside the cryostat. As shown in Figure \ref{Fig:SCtrans}b), almost all power is reflected back on resonance below $T_c$. These methods confirm that the traps are superconducting while ions are trapped in the presence of lasers and rf current drive. When the wire structure is current biased with 1-10~mA less than the critical current, the 405~nm, 422~nm, and 460~nm lasers grazing incident on the trap cause it to transition to the normal state. However, under normal trapping conditions, the lasers have no effect on the measured critical current. 

The ion heating rates in all traps are obtained by measuring the average number of motional quanta with a varied delay after ground state cooling on the S-D optical transition. The number of motional quanta is determined by probing the sidebands of the transition using the shelving technique, and comparing the ratio of shelving probability on each sideband \cite{Turchette:00}. The measured heating rate is weakly dependent on the rf voltage and dc compensation voltages, so these parameters are varied during measurement to find the operating point that gives the lowest heating rate. This value can still vary between different days depending on the trap's processing history, temperature cycling, and other unknown factors; but is typically within the same order of magnitude. The heating rates of Nb and NbN traps are comparable to the lowest heating rates of traps of the same design and tested in the same cryogenic experiment but made with normal metals including Au, Ag, and Al, as listed in Table \ref{Tbl:allheating}.

\begin{table}[tb]
\begin{tabular}{ccc}
\hline\hline
Trap &\quad heating rate (q/s) &\quad $S_E$ ($10^{-15}$ V$^2$/m$^2$/Hz)\\
\hline\hline
NbN  & 16(1)   & 192(12) \\
Nb   & 2.1(3) & 25(4) \\
Nb-g & 4.2(8) & 48(12) \\
\hline
Au$^{\rm a}$  & 2.1(4) & 25(5) \\
Ag$^{\rm b}$ & 2.1(2) & 25(3)\\
Al & 7.0(8) & 84(10)\\
\hline\hline
\end{tabular}
\begin{flushleft}
$^{\rm a}$Ref. \textup{ \cite{Labaziewicz:08b} } \hfill \\
$^{\rm b}$Ref. \textup{ \cite{Labaziewicz:08} } \hfill 
\end{flushleft}
\caption{\label{Tbl:allheating} Heating rate in quanta/second of traps made of superconducting and normal metals measured at cryogenic temperatures. Conversion to electric field noise $S_E$ is scaled to 1~MHz assuming $1/f$ scaling \cite{Labaziewicz:08}.\newline }
\end{table}

\begin{figure}[tb]
\includegraphics[width=3.375in]{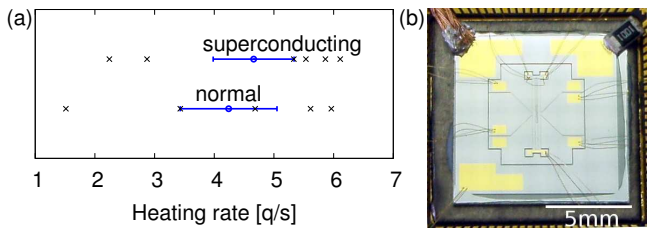}
\caption{\label{Fig:Nbgheating} (a) Heating rate in Nb-g in the normal and superconducting state, with mean and errorbars at 1 standard deviation. Individual data points are shown. The difference between normal and superconducting data is not significant.  (b) Photo of Nb-g. Top right corner: 1~k$\Omega$ resistor for heating the trap; top left corner: copper braid to thermally connect the trap to the helium baseplate. }
\end{figure}

We measured the heating rate above and below the superconducting transition in the Nb-g trap. For the data above $T_c$, 3~mA of current is driven to the 1~k$\Omega$ resistor so as to heat the trap just past $T_c$ as observed by monitoring resistance of the wire structure, corresponding to 9~mW of power dissipated on the trap. In a subsequent cooldown, we mounted RuO$_2$ temperature sensors on the trap\cite{Labaziewicz:08b} and estimate that the operating temperature in the normal state is $\sim$2~K above $T_c$. The trap heating rate is measured immediately before and after this change as shown in Figure \ref{Fig:Nbgheating}. Measurements above and below $T_c$ are interleaved and taken in quick succession, and they are found to be comparable. All data were taken within one cooldown over two days.

The negligible change in heating rate across $T_c$ suggests that buried defects have little effect on anomalous heating. First it is useful to note that at cryogenic temperatures, the expected level of Johnson noise is on the order of 10$^{-20}$ V$^2$/m$^2$/Hz, while the field noise as measured by the ion is on the order of 10$^{-14}$ V$^2$/m$^2$/Hz. Thus it is not surprising that removing the Johnson noise may not have much effect on anomalous heating. The remaining explanation is that anomalous heating is predominantly a surface effect and is unrelated to resistivity. The distinction between surface and bulk is given by the London penetration depth, which in Nb is about an order of magnitude less than the 400~nm film thickness. The results here are still consistent with the current theory of patch potentials on metal surfaces. For superconductors, a recent theory proposed that surface plasmons can be an additional source of electromagnetic noise \cite{Henkel:08}.

In one Nb trap we tested, the heating rate was measured on multiple instances over the period of over one year. During this time the trap was installed and removed from the cryostat multiple times and exposed to air in between with no processing or cleaning. The lowest heating rate obtained during any data run shows no significant variation over the year. In contrast, in many of our electroplated gold traps the heating rate can change by an order of magnitude between temperature cyclings, and after a few months in storage, increase in surface roughness and color changes along all electrode edges observable under an optical microscope is apparent. 


In conclusion, we have demonstrated superconducting ion traps that show good trapping stability and low heating rates. The heating rate does not change appreciably across $T_c$, indicating that anomalous heating is primarily a surface effect unrelated to bulk resistivity. Though the anomalous heating was not reduced by superconductivity, the consistency of the low anomalous heating through temperature cycling and exposure to air is an advantage over other materials such as electroplated gold traps. The feasibility of superconducting ion traps invite the possibility of integrating trapped ions with superconducting devices such as Josephson junctions and SQUIDs, though the compatibility of such devices is open to investigation. Recent progress in using the ion as an extremely sensitive detector of forces and charges \cite{Maiwald:09, Biercuk:10, Harlander:10} also suggest the possibility of detecting superconducting vortices with trapped ions. Magnetic flux trapped in vortices would modify the magnetic field above the superconductor. The vortex density is determined by the applied external field during cooling across the superconducting transition. The resulting change in local magnetic field can be detected by the ion via the Ramsey method on a narrow transition. An estimate of the ion's sensitivity to magnetic field\cite{Maiwald:09} of 1.1$\times 10^{-11}$ T/$\sqrt{\tau/{\rm Hz}}$ and typical ion height of 100~$\mu$m are comparable to parameters in early experiments that demonstrated vortex detection using SQUIDs \cite{Minami:91, Mathai:92}. Such coupling to superconducting vortices have been demonstrated with trapped neutral atoms in a recent experiment\cite{Muller:10}.

We thank Adam McCaughan for assistance in trap fabrication. This work was supported by the COMMIT Program with funding from IARPA, the DARPA Quest program, and the NSF Center for Ultracold Atoms.

%

\end{document}